%% file: main.tex
\documentclass[a4paper, amsfonts, amssymb, amsmath, reprint, showkeys, nofootinbib, twoside,aps,pra]{revtex4-2}
\usepackage[english]{babel}
\usepackage[utf8]{inputenc}
\usepackage[colorinlistoftodos, color=green!40, prependcaption]{todonotes}
\input{preamble}
\usepackage[pdftex, pdftitle={Article}, pdfauthor={Author}]{hyperref} 
\usepackage{graphicx}
\usepackage{subcaption}
\usepackage{tikz}
\usetikzlibrary{calc}

\begin{document}
\title{Improve Variational Quantum Eigensolver by Many-Body Localization}

\author{Xin Li}
    \email{lixinphy@bit.edu.cn}
    \affiliation{Center for Quantum Technology Research and Key Laboratory of Advanced Optoelectronic Quantum Architecture and Measurements (MOE), \\ School of Physics, Beijing Institute of Technology, Beijing 100081, China}

\author{Zhang-qi Yin}
    \email{zqyin@bit.edu.cn}
    \affiliation{Center for Quantum Technology Research and Key Laboratory of Advanced Optoelectronic Quantum Architecture and Measurements (MOE), \\ School of Physics, Beijing Institute of Technology, Beijing 100081, China}

\date{\today} 

\begin{abstract}
Variational quantum algorithms have been widely demonstrated in both experimental and theoretical contexts to have extensive applications in quantum simulation, optimization, and machine learning. However, the exponential growth in the dimension of the Hilbert space results in the phenomenon of vanishing parameter gradients in the circuit as the number of qubits and circuit depth increase, known as the barren plateau phenomena. In recent years, research in non-equilibrium statistical physics has led to the discovery of the realization of many-body localization. As a type of floquet system, many-body localized floquet system has phase avoiding thermalization with an extensive parameter space coverage and  have been experimentally demonstrated can produce time crystals.
We applied this circuit to the variational quantum algorithms for the calculation of many-body ground states and studied the variance of gradient for parameter updates under this circuit. We found that this circuit structure can effectively avoid barren plateaus. We also analyzed the entropy growth, information scrambling, and optimizer dynamics of this circuit. Leveraging this characteristic, we designed a new type of variational ansatz, called the 'many-body localization ansatz'. We applied it to solve quantum many-body ground states and examined its circuit properties. Our numerical results show that our ansatz significantly improved the variational quantum algorithm.
\end{abstract}

\maketitle

\section{Introduction } \label{sec:intro}
In recent years, the rapid advancement of quantum computing has opened up new possibilities for tackling complex problems, with the Variational Quantum Eigensolver (VQE) garnering significant attention. VQE is an approximate algorithm designed for noisy intermediate-scale quantum (NISQ) devices\cite{preskillQuantumComputingNISQ2018,RevModPhys.94.015004}. The VQE algorithm aims to tune variational wavefunctions applied on NISQ devices via classical computation to minimize the cost function, thereby solving problems\cite{peruzzo2014variational, mcclean2016theory,tillyVariationalQuantumEigensolver2022}. A series of theoretical and experimental studies have demonstrated the significant application value of this hybrid quantum-classical algorithm in quantum many-body physics\cite{Continentino2021,Ven2020,hensgens2017quantum,TARRUELL2018365}, quantum chemistry\cite{Deglmann2014,WilliamsNoonan2017,Heifetz2020,kandalaHardwareefficientVariationalQuantum2017,RevModPhys.92.015003,kandalaHardwareefficientVariationalQuantum2017,cerezo2021variational}, quantum machine learning\cite{jeswal2019recent}, quantum optimization\cite{moll2018quantum}.  
    
However, because of the noise of quantum device and entanglement between the qubits, all types of variational quantum algorithms have the risk of vanishing gradients, which can occur during training or due to the random initialization\cite{mccleanBarrenPlateausQuantum2018,wangNoiseinducedBarrenPlateaus2021}. This phenomenon refers to the exponential decay of the cost function gradients based on specific optimization properties for a given problem, which is known as the barren plateaus (BPs) problem now. Though akin to the vanishing gradient problem in classical machine learning, the two significant distinctions of the BP problem from the vanishing gradient problem in classical neural networks can have significant implications for variational quantum algorithms. Firstly, the BP problem is contingent on the number of qubits, while the vanishing gradient problem depends on the number of layers. Secondly, further investigations suggest that the BP problem may also be associated with other factors specific to quantum circuits,
including the expressibility of the ansatz, the degree of entanglement of the wavefunction, the non-locality of the wavefunction, or the quantum noise. Besides, people believe that the BP relates to the orthogonality catastrophe in quantum many-body physics\cite{tillyVariationalQuantumEigensolver2022, mccleanBarrenPlateausQuantum2018}. 

After Ref.~\cite{mccleanBarrenPlateausQuantum2018}, there have been many papers studied the BP problem. Ref.~\cite{wangNoiseinducedBarrenPlateaus2021}  conducted a deeper study on the differences between entanglement-induced BP and noisy-induced BP. Various suggestions have been proposed to alleviate BP, such as designing various initialization circuits to avoid BP\cite{grantInitializationStrategyAddressing2019,holmesConnectingAnsatzExpressibility2022,sackAvoidingBarrenPlateaus2022,wiersemaExploringEntanglementOptimization2020}. However, these circuits are either shallow or subject to various constraints. Other papers have studied the deeper relationship between BP and circuit structures, such as reducing BP by controlling entanglement and combining classical shadow methods to avoid BP during parameter tuning, but such methods are overly complicated.

Meanwhile, recent research on non-equilibrium system have indicated that the dynamical evolution of non-equilibrium systems may lead to phases distinct from those of equilibrium systems, such as dynamical phase transitions Many-Body Localization (MBL)\cite{abanin2019colloquium}, 
Quantum Many-Body Scars, and discrete time crystals (DTCs)\cite{khemani2019brief,rovnyObservationDiscreteTimeCrystalSignatures2018}. Research indicates that double periodic Floquet evolutions can realize various discrete time crystals, such as prethermal time crystals\cite{yingFloquetPrethermalPhase2022,palTemporalOrderPeriodically2018} and MBL time crystals\cite{ippolitiManyBodyPhysicsNISQ2021}. Compared to prethermal DTCs, which require complex many-body interactions and may fail over long timescales, MBL DTCs have simpler time evolution operators and can persist for longer durations. Utilizing MBL configurations combined with double periodic Floquet evolutions\cite{ippolitiManyBodyPhysicsNISQ2021,Zhang_2016}, Ref~\cite{miTimecrystallineEigenstateOrder2022} extensively discusses a scheme to realize MBL-DTC in superconducting quantum device, noting that this MBL-DTC configuration occupies a wide range of MBL phases in parameter space. followed by a series of experiments in ion traps and other platforms that achieved similar DTC configurations\cite{liuDiscreteTimeCrystal2023}, these research indicate that Floquet MBL can be stably and efficiently simulated in existing 
different NISQ devices, maintaining long-time existence.
Inspired by these studies, from the perspective of circuit structure, the VQE parameter circuit can also be viewed as a type of unitary parameterized circuit. We integrate methods for studying the evolution of non-equilibrium states into the investigation of the structure of VQE circuits. 

In this article, we unveil the relationship between entanglement-induced BPs and thermalization through numerical simulations. We demonstrate that the circuit structure of MBL effectively avoids BPs. Moreover, we find that initializing the parameters of the VQE circuit in the MBL regime effectively circumvents BPs. This initialization strategy exhibits a certain level of robustness against noise and other perturbations, making it relatively effective compared to general VQE circuits and easy to implement in NISQ device. Based on these properties, we designate this initial configuration as a typical variant of HE ansatz called "MBL-ansatz". 

The rest of paper is organized as follows.
In Section~\ref{Sec2}, we will first introduce the variational ansatz and the barren plateau problem, than we will explain the details of the MBL model we used, elaborate on the parameter distribution of our circuits, and how we selected them. In section~\ref{Sec3}, we will introduce our main resultes: by comparing with conventional random parameter circuits, we found that circuits with an MBL structure can avoid BPs even in very deep circuits, while circuits in the thermal phase do not exhibit such favorable properties. Additionally, by calculating information scrambling and entropy growth with increasing depth, we elucidate the differences in circuit structures under different phases and highlight the excellent characteristics of the MBL structure. In Section~\ref{optimizer}, we performed a complete gradient optimization of our designed ansatz, obtaining the dynamics of the optimization iteration process, demonstrating the superiority of our ansatz in handling many-body ground state problems.

\section{Method} \label{Sec2}

\subsection{Variational Ansatz and Barren Plateaus} \label{habp}
The core of the Variational Quantum Algorithms(VQAs) lies in utilizing quantum circuits to prepare parameterized quantum states on NISQ devices, constructing cost functions with parameters, and using classical computers to optimize these cost functions to obtain their minimum values\cite{kandalaHardwareefficientVariationalQuantum2017}. In the context of VQAs, the phenomenon of BPs manifests as a drastic reduction in the variance of parameter gradients as the number of qubits increases\cite{mccleanBarrenPlateausQuantum2018}.

A $N$-qubit, $D$-depth random parameterized quantum circuits(RPQCs) for VQAs usually has form
\begin{equation}\label{rpqcs}
U(\boldsymbol{\theta}) = \prod_{d=1}^{D} U^d(\boldsymbol \theta^d) = \prod_{d=1}^{D} \prod_{k=1}^{N} e^{-\theta(d,k)\text{i} \Omega_k} 
\end{equation}
where $\{\Omega_1,\dots,\Omega_N\}$ are hermitian Pauli-string generators, thus $e^{-\theta(l,k)i \Omega_k}$ is rotation gate. A periodic ansatz consists of this randomized unitary circuit, an initial state$\rho_0 = U_0|0\rangle\langle0|U_0^\dagger$, here $|0\rangle=|0\rangle^{\otimes N}$ is the zero product state, and a Hermitian operator $H$. The $H$ usually a many-body hamiltonian. The cost function of a VQE is the expectation value 
\begin{equation}\label{cost}
    E(\boldsymbol{\theta}) = \text{Tr}[U(\boldsymbol\theta)\rho_0 U^\dagger(\boldsymbol\theta)H]
\end{equation}

In the VQE algorithm, the phenomenon of BPs can be formally described as the exponential decrease in the variance of parameter gradients with an increase in the number of qubits.
If the cost function Eq.~(\ref{cost}) exhibits a BP, for any $\theta_i \in \boldsymbol{\theta}$:
\begin{equation}
\operatorname{Var}_{\boldsymbol{\theta}\sim\nu}\left[\partial_{\theta_i} E(\boldsymbol{\theta})\right] \sim \mathcal{O}\left(\frac{1}{b^{N}}\right)
\end{equation}
with $b > 1$. $\nu$ means a distribution of the parameters, usually is the uniform distribution in a range. The uniform distribution over parameters means the unitary $U(\boldsymbol{\theta})$ sampled from a ensemble $\boldsymbol{U}$ via Haar measure and forms an unitary 2-design\cite{mccleanBarrenPlateausQuantum2018,sackAvoidingBarrenPlateaus2022,wangNoiseinducedBarrenPlateaus2021}s. 

\begin{figure*}[htbp]
    \centering
    \begin{minipage}{0.3\textwidth}
        \centering
        \includegraphics[width=\textwidth]{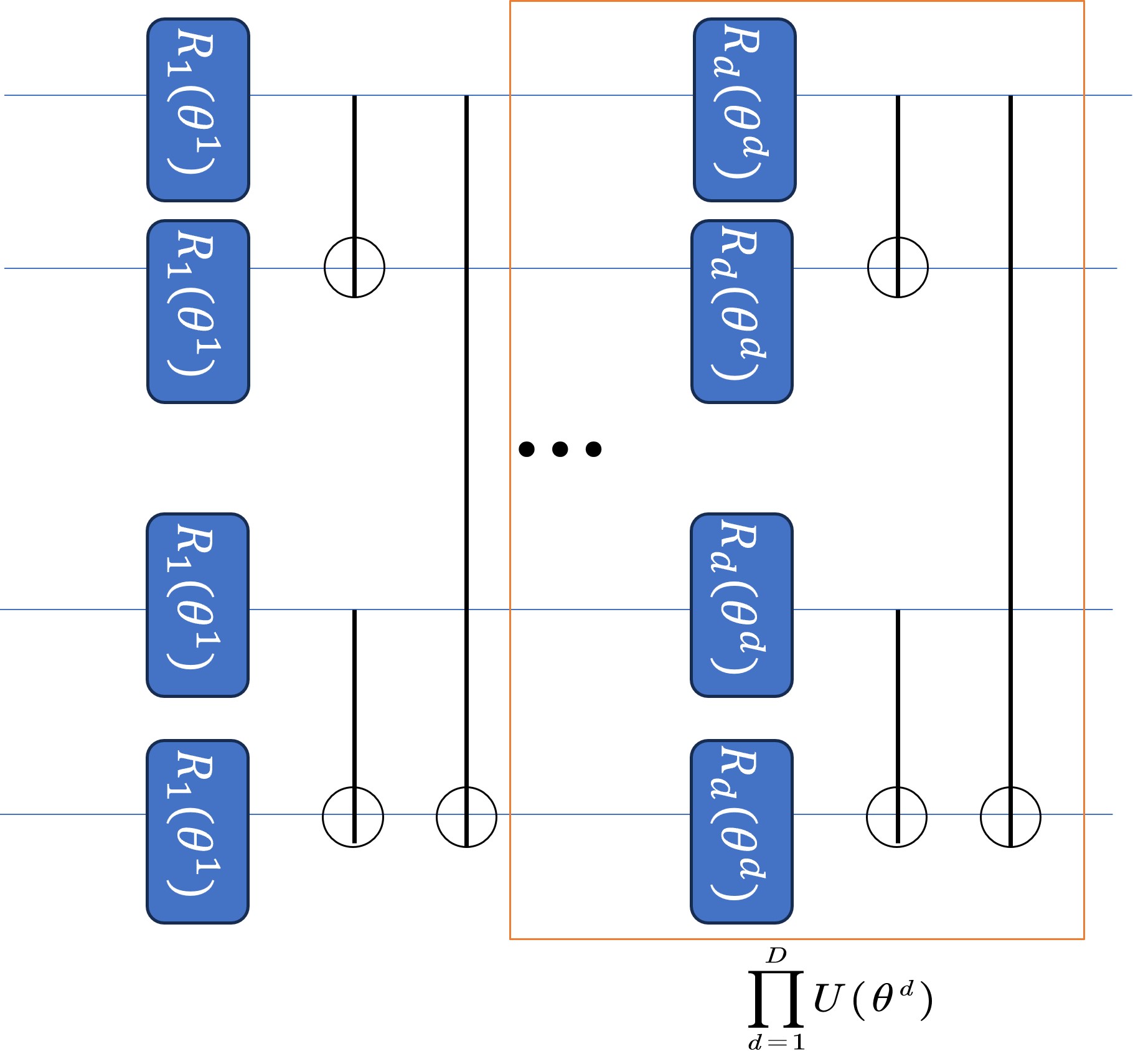}
        \begin{tikzpicture}[remember picture, overlay]
            \node[anchor=north east, font=\footnotesize, inner sep=0pt] at (-3.,5.6) {(a)};
        \end{tikzpicture}
        \label{fig:1a}
    \end{minipage}
    \hfill
    \begin{minipage}{0.3\textwidth}
        \centering
        \includegraphics[width=\textwidth]{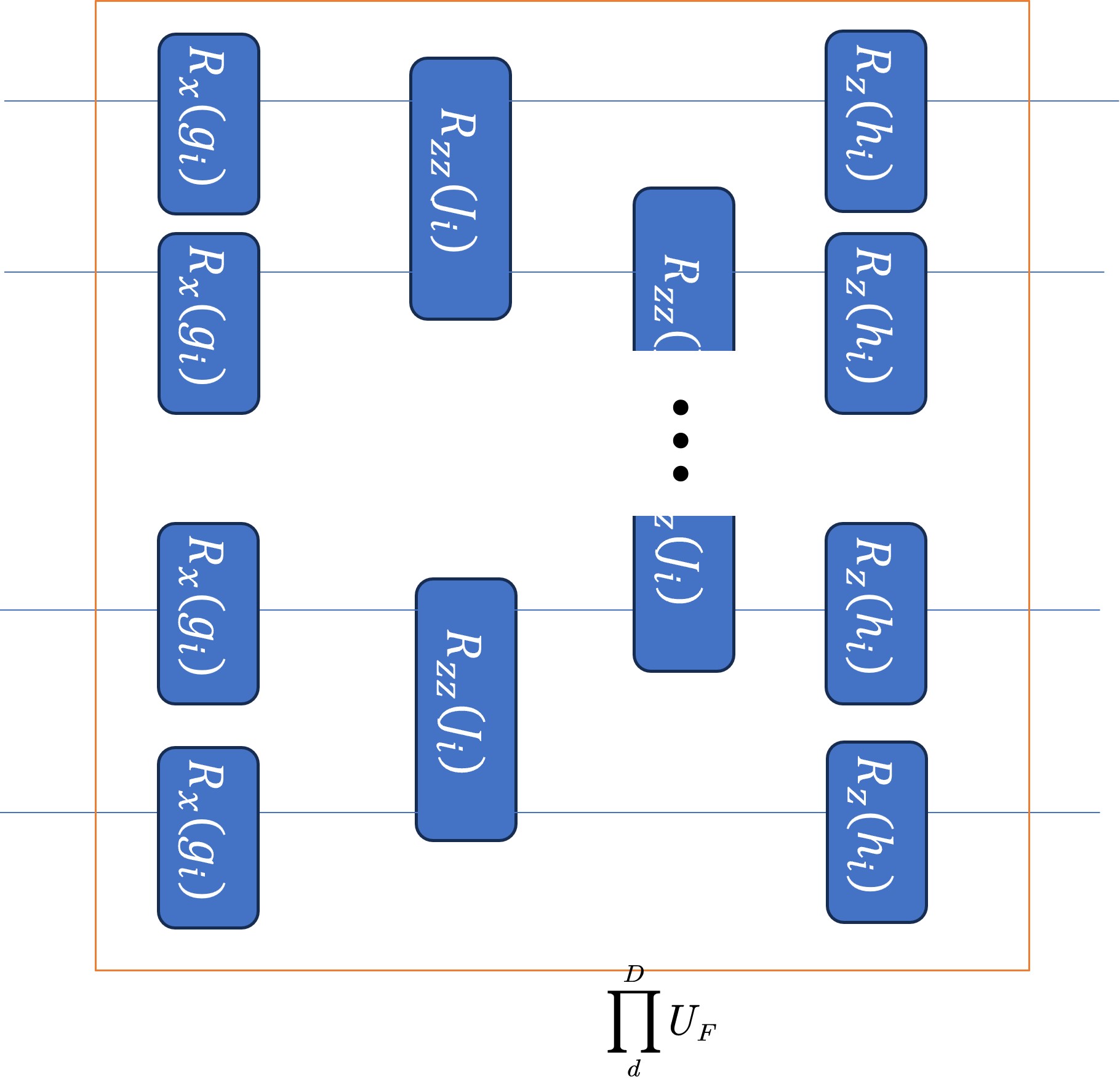}
        \begin{tikzpicture}[remember picture, overlay]
            \node[anchor=north east, font=\footnotesize, inner sep=0pt] at (-3.,5.6) {(b)};
        \end{tikzpicture}
        \label{fig:1b}
    \end{minipage}
    \hfill
    \begin{minipage}{0.3\textwidth}
        \centering
        \includegraphics[width=\textwidth]{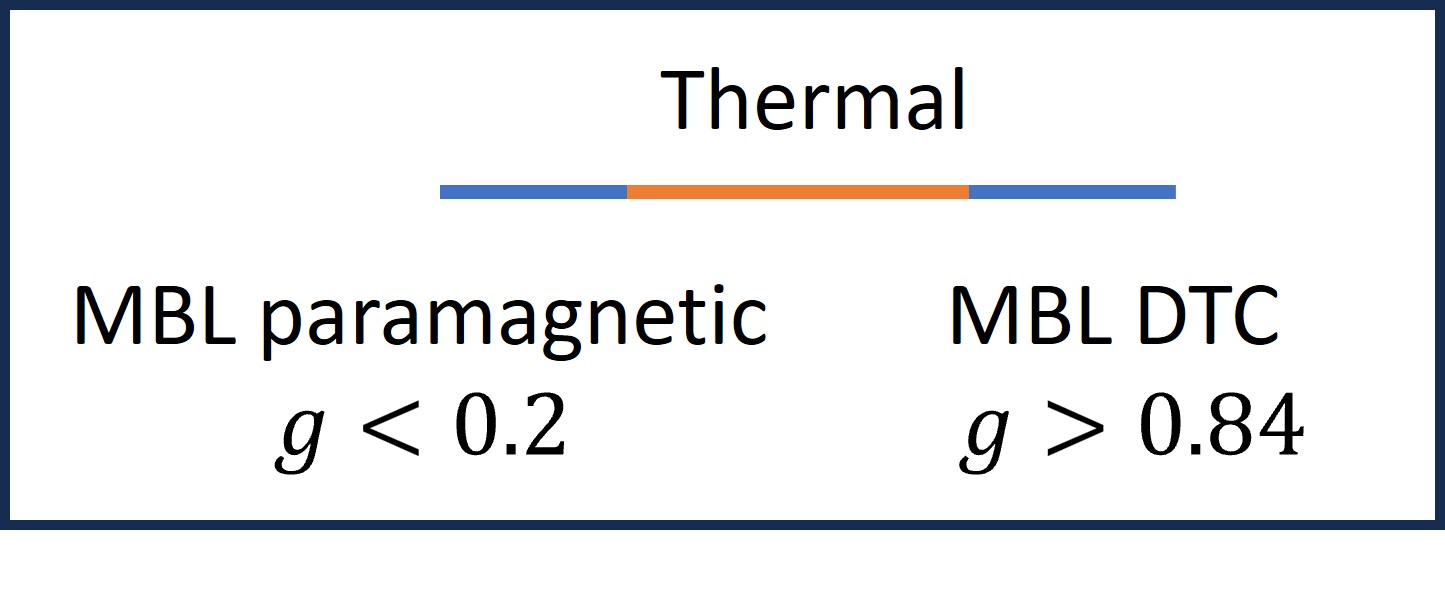}
        \begin{tikzpicture}[remember picture, overlay]
            \node[anchor=north east, font=\footnotesize, inner sep=0pt] at (-3.,3) {(c)};
        \end{tikzpicture}
        \label{fig:1c}
    \end{minipage}

    \caption{(a) Circuit structure for traditional HEA, The unitary is composed of a layer of single-qubit arbitrary rotation gates and an entanglement layer made up of CNOT gates connecting pairs of adjacent qubits. When repeated for $D$ depth, such a circuit is typically thermal, and such a unitary satisfies the unitary 2-design. (b) Our MBL circuit structure consists of single-qubit X rotations, Z rotations, and ZZ rotations between adjacent qubits. By adjusting the range of conditional parameters, this circuit can achieve different phases. If the parameters of the Z layer and ZZ layer are fixed with a random distribution, (c) shows the phase diagram under different parameters of the X layer. Around \( g \approx 0.2\pi \), there is a phase transition from the PM phase to the thermal phase, and around \( g \approx 0.8\pi \), there is a transition from the thermal phase to the discrete time crystal (DTC) phase. In our main result, we select the parameters \( g = 0.1\pi \) and \( g = 0.9\pi \) to demonstrate the results in the PM and DTC phases, respectively.}
    \label{fig1}
\end{figure*}
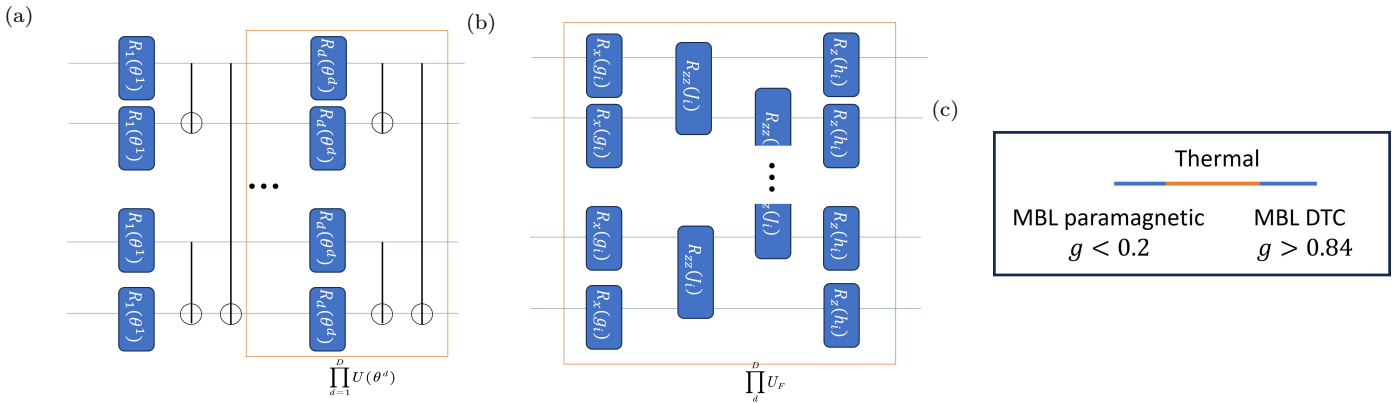

To illustracte the BP, we first introduce the traditional haredware-efficient ansatz(HEA) that has parameterized circuit as\cite{kandalaHardwareefficientVariationalQuantum2017}:
\begin{equation}\label{hea}
    U(\boldsymbol{\theta}) = \prod_{d=1}^{D} W_d \left(\prod_{k=1}^{N} R^k_d(\theta_d^k)\right)
\end{equation}
Where $D$ represents the number of layers or circuit depth for the circuit, and $L$ denotes the number of qubits. $W_d$ consists of all the adjacent qubits' CNOT gates. $R(\theta_k) = \exp(-k \theta_k \hat\tau)$ where $\hat\tau \in \{\sigma^x,\sigma^y, \sigma^z\}$ is single Pauli gate in $k$-th qubit. This RPQC is composed solely of single-qubit random rotations and CNOT gates. Fig. \ref{fig1} (a) depicts the circuit diagram of this ansatz. Not that the CNOT gate can be obtained by fixing the corresponding parameters in certain rotation gates. 

According to the unitary 2-design of the gradient computation, the HEA exhibits the BPs with cost function Eq.~(\ref{cost}), i.e. the parameter gradients will satisfy a exponentially vanish throughout the training trajectory. It make this RPQC untrainable\cite{mccleanBarrenPlateausQuantum2018}.

\subsection{Floquet Many-Body Localization}
In this subsection, we will provide a detailed introduction to the MBL Floquet system and the specifics of the model we use.
MBL is a phenomenon in quantum many-body systems where the system does not reach thermal equilibrium even over long periods, instead remaining in a localized quantum state. This localization is the result of the combined effects of disorder and interactions within the system, leading to the formation of a series of localized integrals of motion, known as \textit{l-bits}\cite{abanin2019colloquium}. These localized integrals of motion are an extensive set of local conservation laws that explain many of the unique features of the MBL phase, including its unusually slow information scramblings\cite{fanOutofTimeOrderCorrelationManyBody2017}, the logarithmic entanglement lightcone\cite{deng2017logarithmic}, the area law entropy growth\cite{bauer2013area,abanin2019colloquium} in any initial state, in another word in every eigenstates even in high energy states, the different of orthogonality catastrophe\cite{deng2015exponential} and the possibility of novel forms of "localization-protected" order\cite{Huse_2013}.

Floquet MBL\cite{ponte2015many} is a special case of MBL involving periodically driven quantum systems. In Floquet systems, the temporal evolution is not generated by a constant Hamiltonian but is described by a discrete time-evolution operator that is repeatedly applied over subsequent time steps. This periodic driving can lead to new physical phenomena, including localization in the absence of energy conservation\cite{morningstarAvalanchesManybodyResonances2022,khemani2019brief}.

There are many ways to construct an Floquet MBL system\cite{khemani2019brief,Zhang_2016,morningstarAvalanchesManybodyResonances2022}. In theory, we can construct a unitary that conforms to MBL by sampling random quantum gates on a qubit chain that comply with the GUE (Gaussian Unitary Ensemble), without the need for specific conservation laws\cite{morningstarAvalanchesManybodyResonances2022}. 

Consider a random time evolution operator composed of a circuit of random unitaries coupling even and odd neighboring spins on a qubit chain in turn. Unitaries constructed from such two-qubit gates can always be decomposed into parameterized circuits of single-qubit and two-qubit gates, thus providing the possibility of constructing them into VQAs.
We begin with a general model of a floquet MBL for Eq.~(\ref{uf}), it is an XY-Model periodically "Kicked" by a rotation X-gate. The unitary is:
\begin{equation}
U(T)=\exp \left[-i g \sum_j \sigma_j^x\right] \exp \left[-i T H_{\mathrm{int}}\right]
\end{equation}
Where the "interacton term" $H_{\mathrm{int}}$ is
\begin{equation}
   H_{\mathrm{int}}=\sum_i J_j \sigma_j^z \sigma_{j+1}^z+J_j^{xy} (\sigma_j^x\sigma_{j+1}^x +\sigma_j^y\sigma_{j+1}^y)+h_j^z \sigma_j^z. 
\end{equation}

In such a model, the disorder strength of the parameter \( h^z_j \) (that is, randomness, in VQA) controls whether the unitary is in the MBL regime. The parameter \( J^{xy} \) controls the magnitude of ``hopping,'' which interferes with the formation of MBL. The parameter \( J^Z_j \) determines from a fermionic perspective that this is a many-body interaction, and its disorder will determine the stability of the MBL regime; the greater the disorder strength, the more stable it is. 

As an example for MBL circuit, we simply sets \( J^{xy} \) to zero in this model, with the aim of simplifying the circuit and making it easier to implement on NISQ devices. We only need to set the period \( T \) to \( \frac{1}{2} \), and these parameters then take the form commonly used in VQA circuits. This model can consider as a disorderd Ising Hamiltonian with on-site longitudinal field, $H_c = \sum_{i,i+1}2J_i\sigma^z_i\sigma^z_{i+1}+\sum_i h_i Z_i$ with imperfect qubit flip $H_f = \sum_i g_i \sigma^x_i
$ specify to dynamics
consisting of alternating applications of $H_c$ and $H_f$ for time
$T = \frac{1}{2}$, thus the unitary Operator is
\begin{equation}\label{uf} 
U_F = \exp(-\text{i}\frac{1}{2}H_f)\exp(-\text{i}\frac{1}{2}H_c)
\end{equation}
It can be seen that MBL can be stably maintained over a wide parameter space, which makes it possible for us to perform stable gradient updates on the parameters. The most important two-qubit ZZ-rotation can even cover the entire range of available parameters, which also makes our circuit somewhat resistant to noise. There has already been some discussion of this in earlier theoretical research\cite{miTimecrystallineEigenstateOrder2022,ippolitiManyBodyPhysicsNISQ2021}. In earlier theoretical research\cite{ippolitiManyBodyPhysicsNISQ2021}, it was used to construct discrete time symmetry and double periodicity to create time crystals. Adjusting the size of \( g \) can make the system in the PM phase, DTC phase, or Thermal phase. We retain \( g \) in our circuit to increase non-commutativity, enriching the rotation of the circuit, and allowing each qubit to rotate not only within the z basis. At the same time, we can easily distinguish the variance and entropy dynamics of each regiem by only adjusting \( g \).

According to previous work\cite{miTimecrystallineEigenstateOrder2022}, to ensure that the evolution conforms to MBL, we need to set the parameters sampled randomly $J_i \in [-1.5\pi,-0.5\pi]$, $ h_i \in [-\pi,\pi]$,if we repeat this $U_F$ in $D$ times with $g_i \in (0,0.2\pi]$ or $g_i \in [0.84\pi,\pi)$, The system will undergo Floquet MBL evolution, particularly, if $g_i \in [0,0.2\pi)$, it will be in the paramagnetic(PM) phase,
$g_i \in [0.84\pi,\pi)$, it will be in the DTC phase\cite{miTimecrystallineEigenstateOrder2022}.
if we repeat $U_F$ D times and chose this circuit as RPQC, thus the parameters be ${J_i}$ ${h_i}$ and $g_i$. We illustrate the gate sequence of this RPQC in Fig.~\ref{fig1} (b).

\section{Result}\label{Sec3}
\subsection{Avoding Barren Plateaus by Many-Body Localization}
We chose the Hermitian operator $H$ in Eq.~(\ref{cost}) as a N-site Heisenberg Model's Hamilatonian:
\begin{equation}\label{xxz}
    \hat{H}_{xxz} = -\sum_{i=1}^{N-1} J(\sigma^x_i\sigma^x_{i+1}+\sigma^y_i\sigma^y_{i+1}) + \Delta\sum_{i=1}^{N-1}\sigma^z_i\sigma^z_{i+1},
\end{equation}
 The Heisenberg model, developed by Werner Heisenberg, one of the models used in statistical physics to model ferromagnetism, Through the Jordan-Wigner transformation, it can also be used to explain strongly correlated fermionic systems and other phenomena. 
It leads the cost funtion Eq.~(\ref{cost}) BP in unitary chaotic circuit like HEA ansatz.

\begin{figure*}[htbp]
    \centering

    \begin{minipage}{0.49\textwidth}
        \centering
        \includegraphics[width=\textwidth]{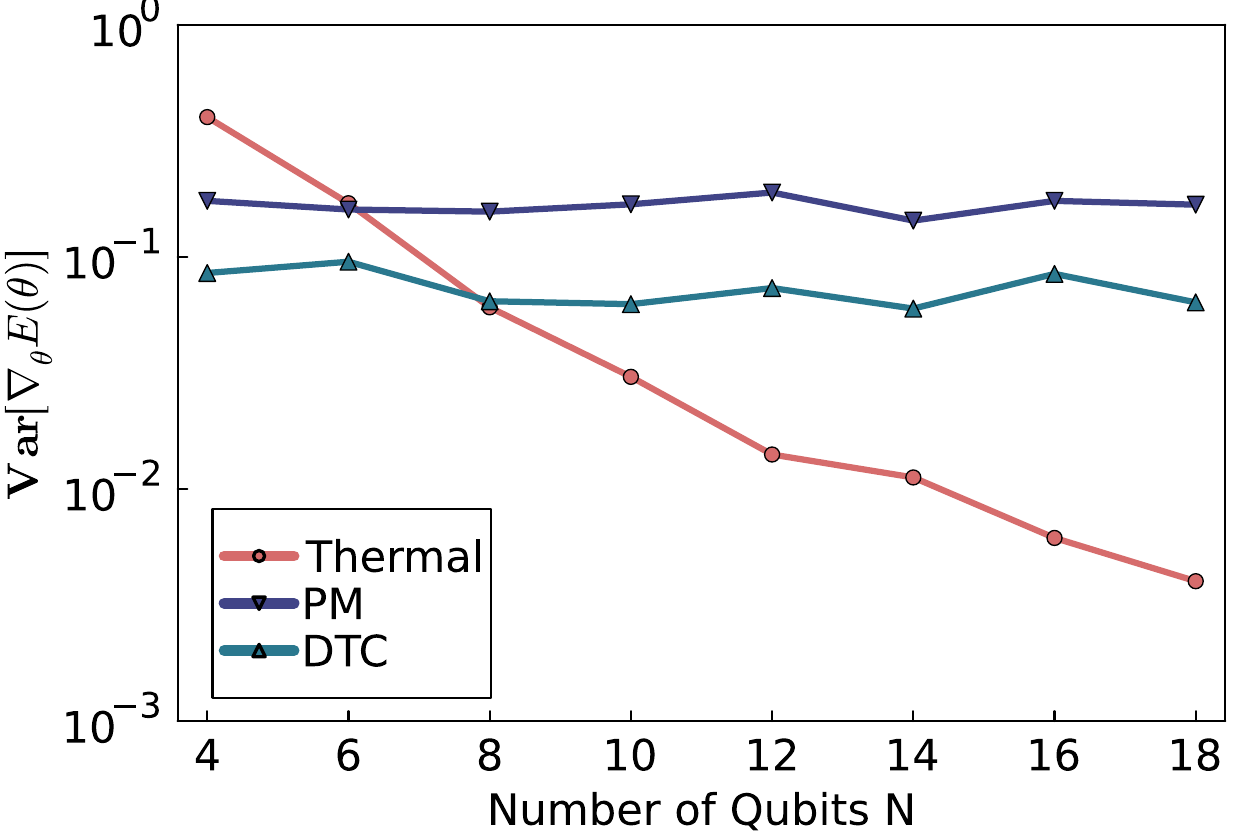}
        \begin{tikzpicture}[remember picture, overlay]
            \node[anchor=north east, font=\footnotesize, inner sep=0pt] at (3.6,5.6) {(a)};
        \end{tikzpicture}
    \end{minipage}
    \hfill
    \begin{minipage}{0.49\textwidth}
        \centering
        \includegraphics[width=\textwidth]{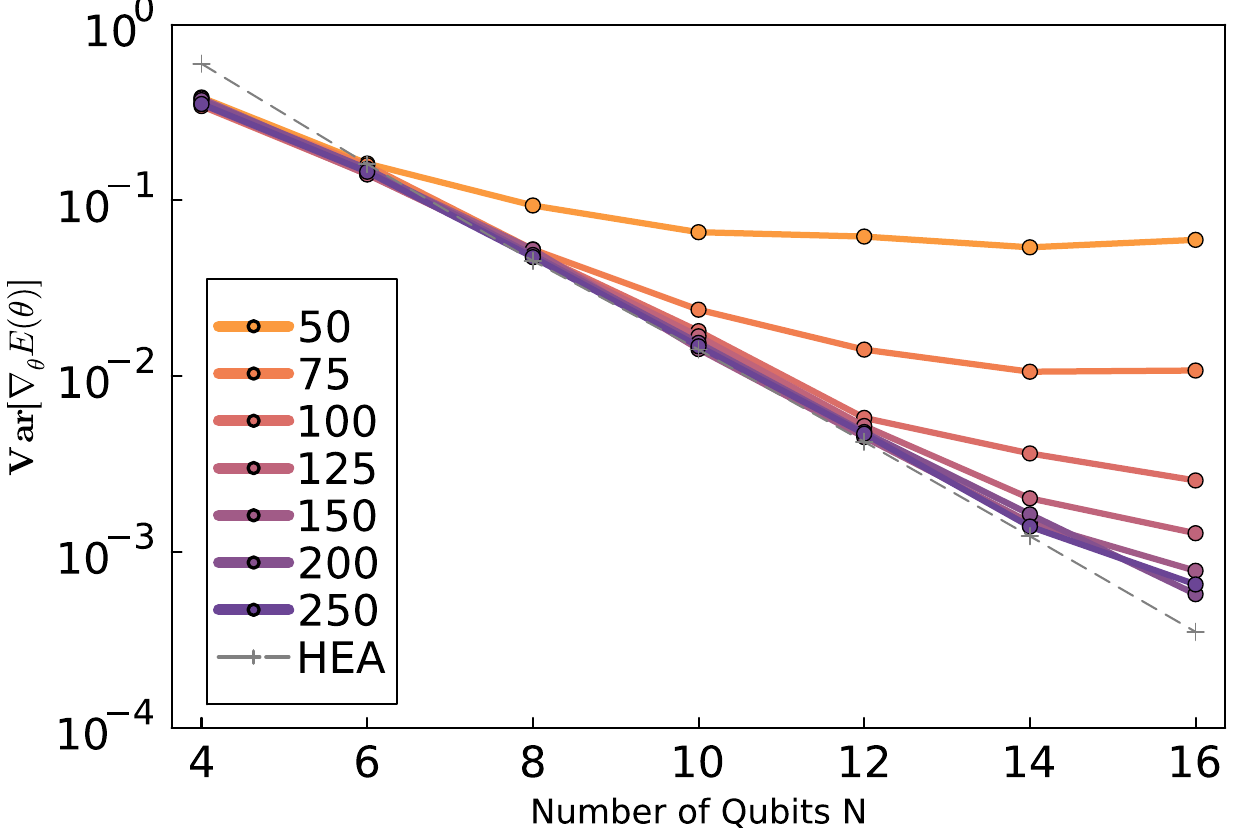}
        \begin{tikzpicture}[remember picture, overlay]
            \node[anchor=north east, font=\footnotesize, inner sep=0pt] at (3.6,5.6) {(b)};
        \end{tikzpicture}
    \end{minipage}

    \vspace{10pt} 
    \begin{minipage}{0.49\textwidth}
        \centering
        \includegraphics[width=\textwidth]{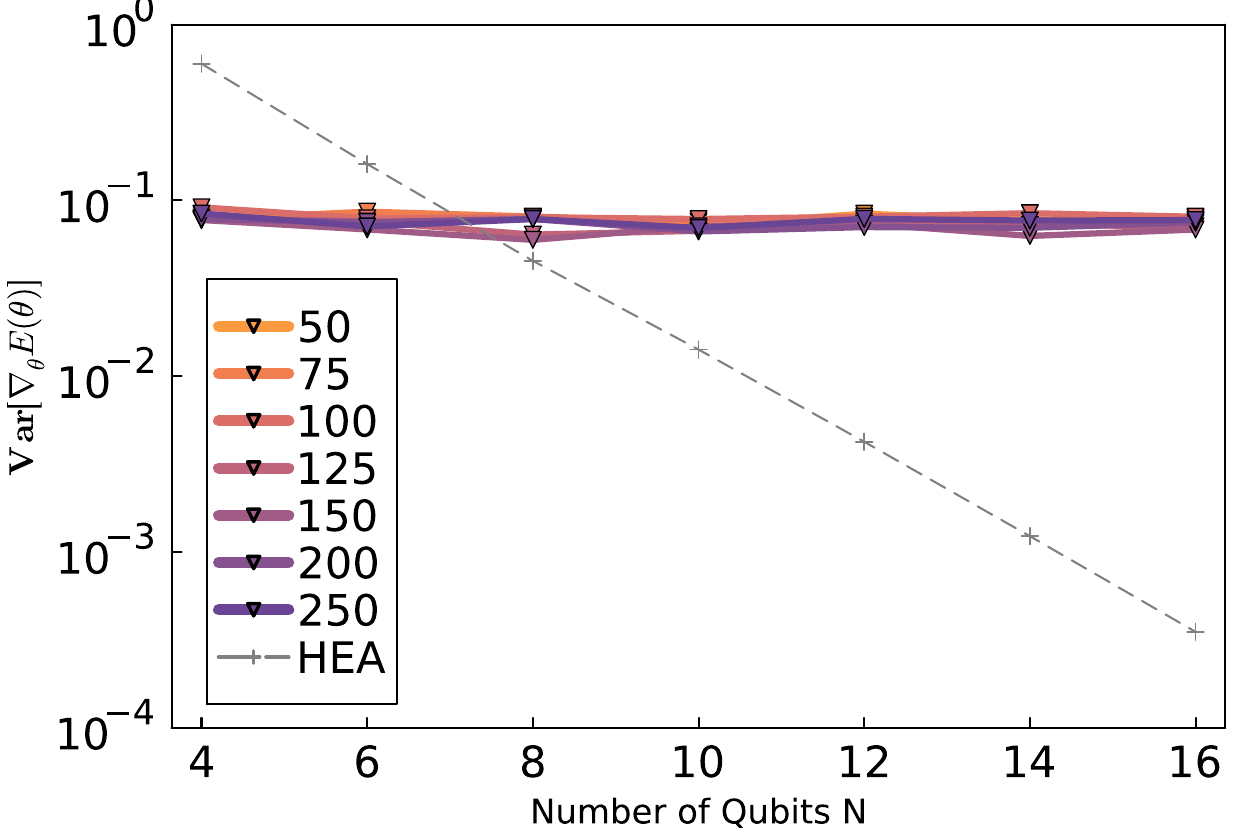}
        \begin{tikzpicture}[remember picture, overlay]
            \node[anchor=north east, font=\footnotesize, inner sep=0pt] at (3.6,5.6) {(c)};
        \end{tikzpicture}
    \end{minipage}
    \hfill
    \begin{minipage}{0.49\textwidth}
        \centering
        \includegraphics[width=\textwidth]{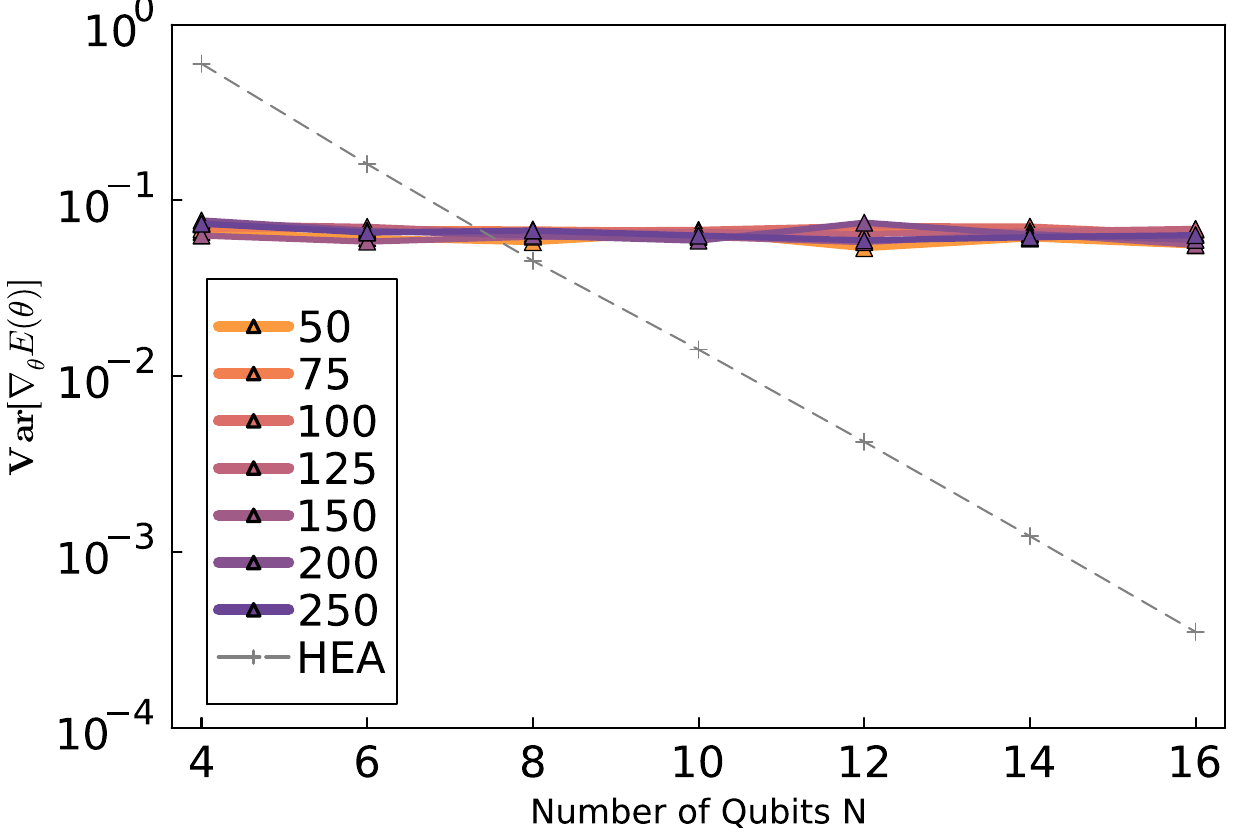}
        \begin{tikzpicture}[remember picture, overlay]
            \node[anchor=north east, font=\footnotesize, inner sep=0pt] at (3.6,5.6) {(d)};
        \end{tikzpicture}
    \end{minipage}
    \caption{(a) The variance of the gradient in different regimes in shallow depth. The circuit depth $D=2\times N$, When the parameter $g$ is chosen within the PM and DTC regions, the variance of the gradient remains constant as the number of qubits increases; in the thermal region, the variance decays as the number of qubits increases, it can be observed that at shallow depths, the decay is smaller than the exponential scale. (b) (c) (d) The variance of the gradient in (b) Thermal ($g=0.5\pi$), (c) PM ($g=0.1\pi$), (d) DTC ($g=0.9\pi$) phase in different depth. The deeper the color, the greater the depth of the circuit.In the thermal region, as the circuit depth increases, the variance of the gradient exhibits the standard exponential decay, and BPs begin to emerge. However, in the PM and DTC regions, even with very deep circuit depths, the variance remains constant at a large value. As a benchmark, the gray dashed line represents the variance of the gradient for the HEA circuit in Fig. \ref{fig1}(a).}
    \label{fig:mainresult}
\end{figure*}
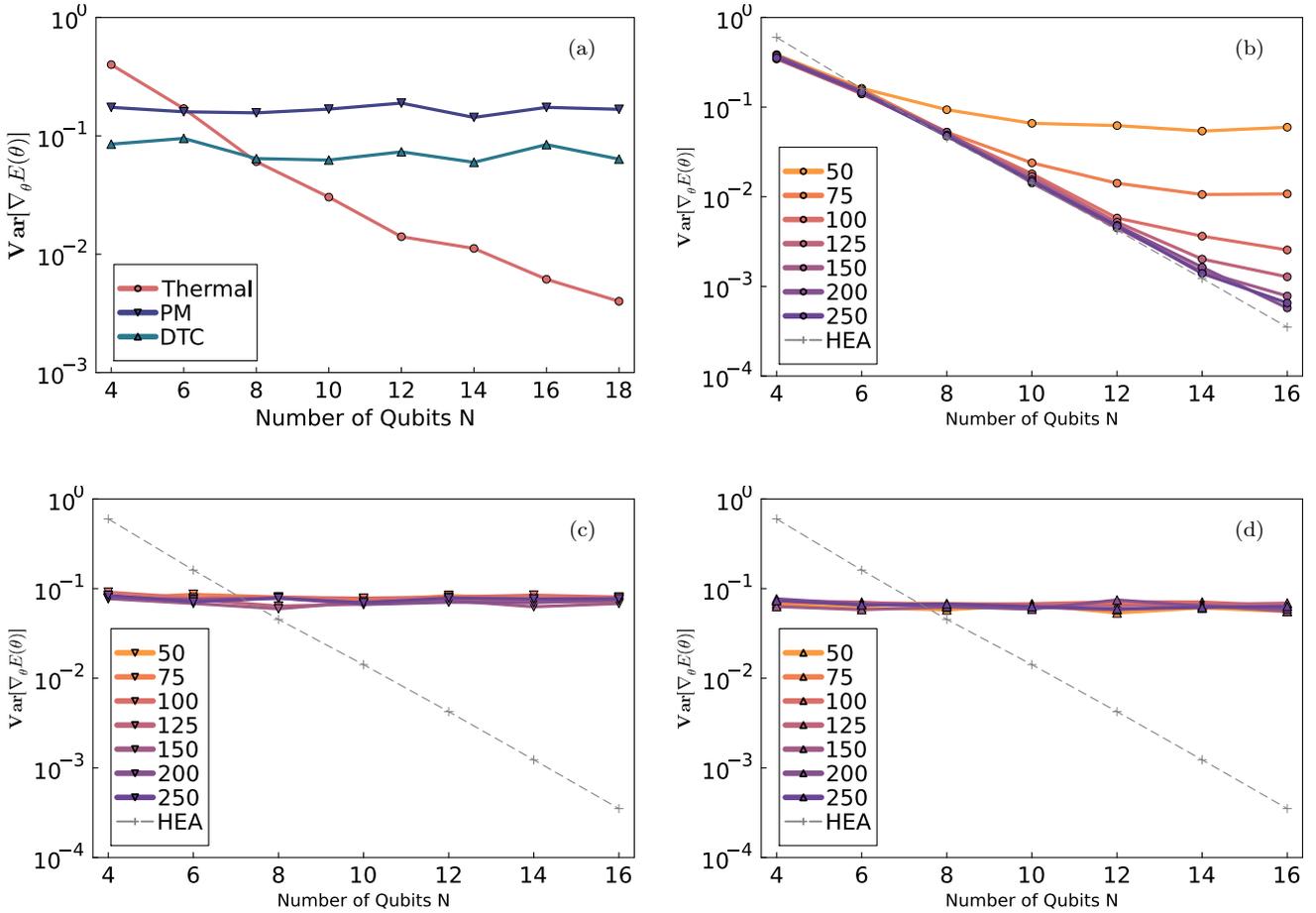

Fig.~\ref{fig:mainresult} shows our main result. As the parameter $g$ vary, the circuit will be in different regions. In Fig.~\ref{fig:mainresult}, to illustrate the issue, we chose $g= 0.1\pi$ as PM circuit, $g=0.9\pi$ as DTC circuit, $g = 0.5\pi$ as thermal circuit, and we the traditional HEA circuit Eq.~(\ref{hea}) Without loss of generality, we have plotted the variance of the gradient of the first $ZZ$-rotation parameter of the circuit i.e. the gradients with respect to the first $J$  parameter in the first MBL circuit block. The same applies to the other parameters. We numerically compute the variance of parameter gradients in different regions. In Fig.~\ref{fig:mainresult} (a), we set the circuit depth to be twice the number of qubits. We can see that, at this scale, the variance of the gradient in the MBL region under both the PM and DTC phases does not increase with the number of qubits. In fact, this characteristic persists even at very deep circuit depths. In Fig.~\ref{fig:mainresult} (b), (c), (d) we calculate the variance of the gradient at different circuit depths in the thermal, PM and DTC regions, respectively. In Fig.~\ref{fig:mainresult} (b) we can see that in the thermal region, although the variance of the gradient does not increase exponentially with the number of qubits in shallow circuits, it still exhibits exponential decay as the depth increases. However, in Fig.~\ref{fig:mainresult} (c) and (d) in the PM and DTC regions, the variance of the gradient remains constant regardless of the circuit depth. In both cases, within the MBL region, the variance of the gradient does not decay exponentially with increasing particle number. 

We will observe that when the parameters are chosen in the MBL regime, thus the circuit leads to entanglment area low rather than volume law for thermal circuit,
the variance of the gradients will not vanish 
exponentially in section~\ref{otoc}. 
It is worth noting that even if the circuit itself is not Floquet, as long as the parameters of $U_F$ are in the MBL region,all maintain the area law, the variance of parameter gradients will not exponentially vanish.

\subsection{Information scrambling}\label{otoc}

\begin{figure}[htbp]
    \centering
    \includegraphics[width=0.49\textwidth]{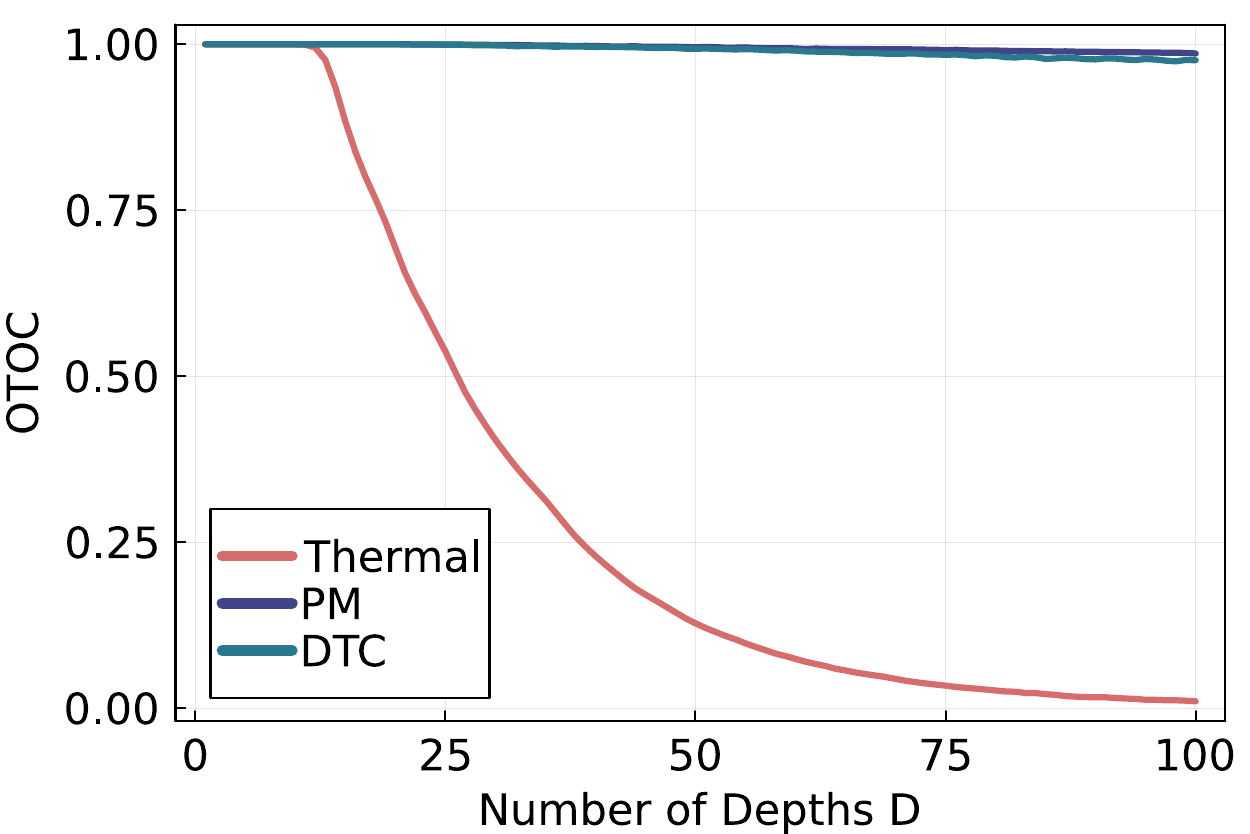}
    \caption{Information scrambing in Diffrent region by caculate the OTOC, We selected \( g = 0.7 \) as the thermal region, \( g = 0.16 \) (PM) and \( g = 0.86 \) (DTC) as the MBL region. We investigated the Out-Of-Time-Ordered Correlator (OTOC) for the X gate at the first and last lattice points. For each case, we calculated the average results from 100 random instances. We can observe that in the shallow circuit, the circuit in the MBL region hardly exhibits any information scrambling, whereas in the thermal region, information scrambling decays rapidly, even beginning to decay to zero at very shallow circuit depths.}
    \label{fig:otoc}
\end{figure}

\begin{figure*}[htbp]
    \begin{minipage}[t]{0.49\textwidth}
        \centering
        \includegraphics[width=\textwidth]{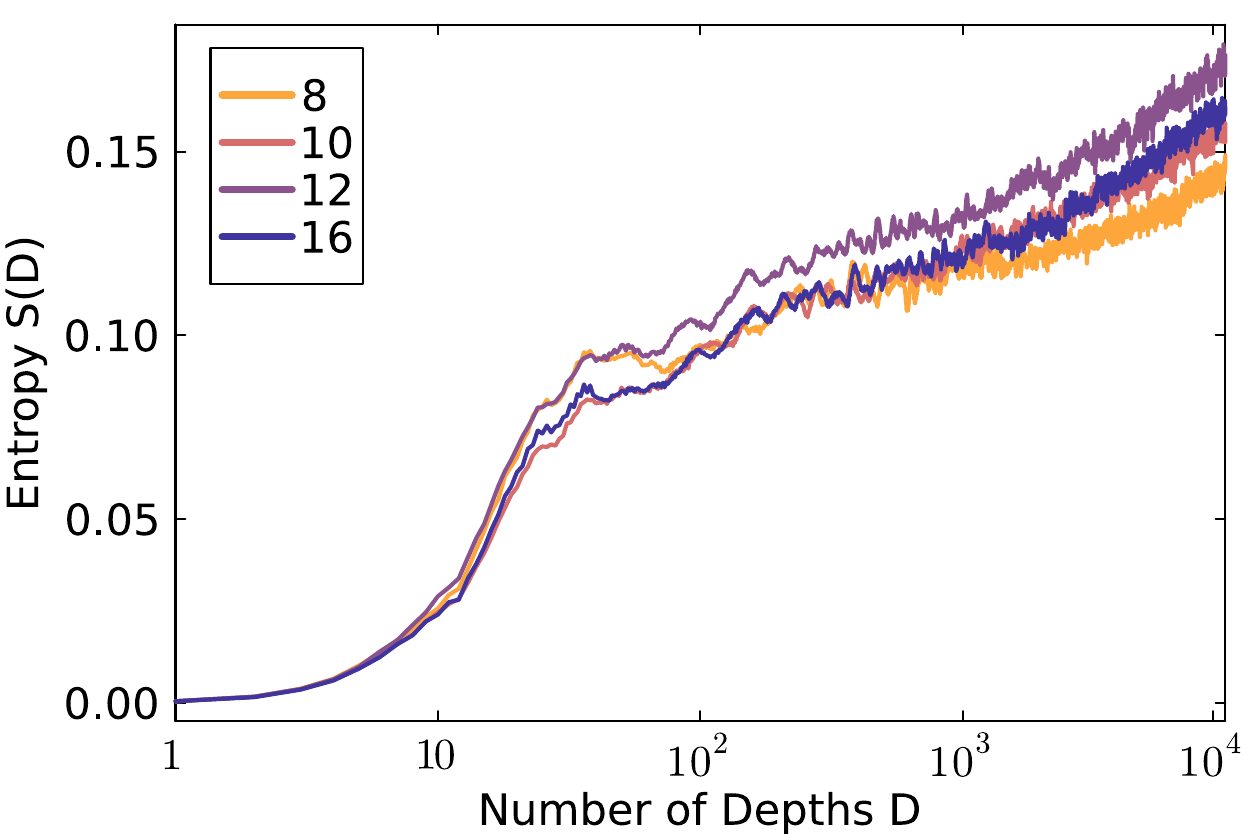}%
        \begin{tikzpicture}[remember picture, overlay]
            \node[anchor=north east, font=\footnotesize, inner sep=0pt] at (-0.85,2.4) {(a)};
        \end{tikzpicture}
    \end{minipage}
\begin{minipage}[t]{0.49\textwidth}
    \centering
    \includegraphics[width=\textwidth]{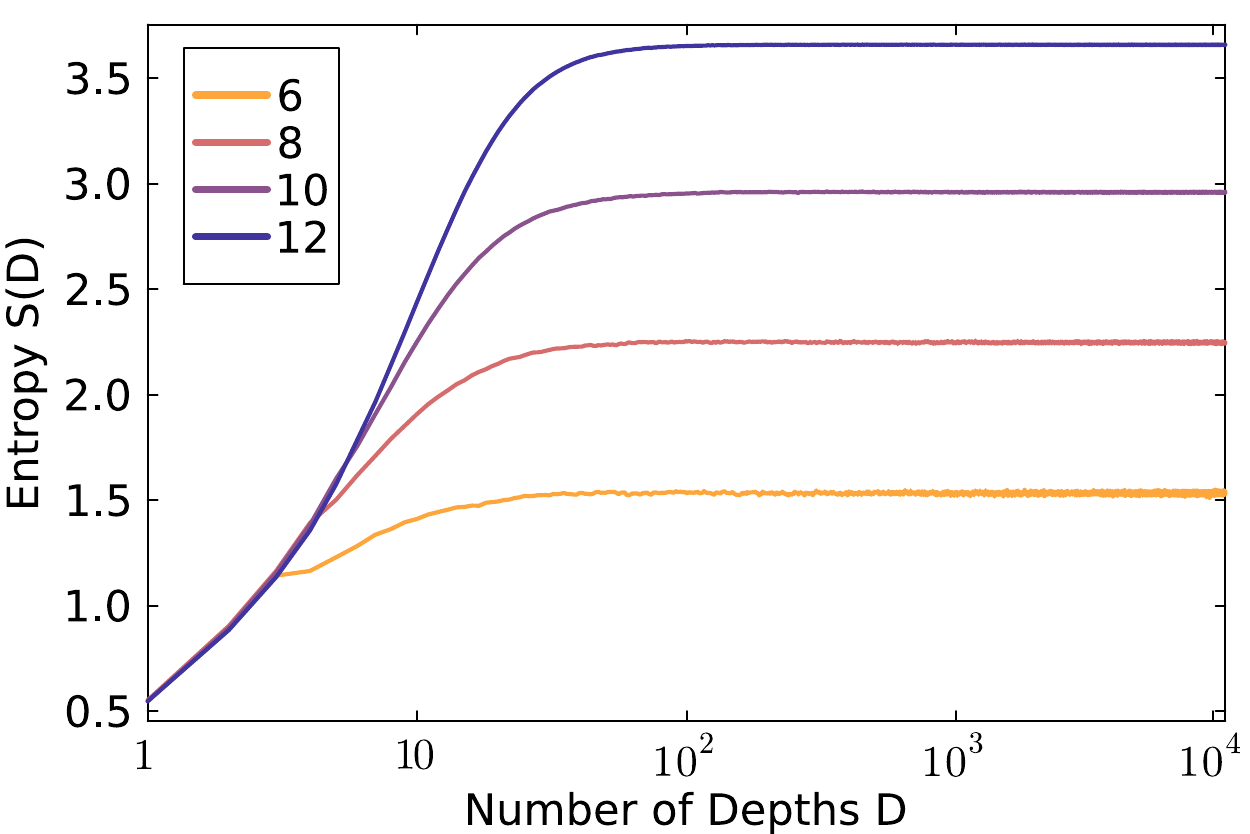}
    \begin{tikzpicture}[remember picture, overlay]
            \node[anchor=north east, font=\footnotesize, inner sep=0pt] at (3.2,2.6) {(b)};
        \end{tikzpicture}
\end{minipage}
\caption{(a) Entropy growth in MBL region, we chose $g=0.9$ and numerical simulate the number of qubits as $N = 8, 10 ,12, 16$, in each case we set 1000 instance and get the averge entropy growth. We can see that the entropy in MBL-DTC region is area law. The situation is the same in the PM phase. (b) Entropy growth in thermal region, we chose $g=0.5$ and mumerical simulate the number of qubits as $N = 6, 8, 10, 12$, in each case we set 1000 instance and get the averge entropy growth, in thermal region, the entropy is volume law.}
\label{entropygrowth}
\end{figure*}

Some previous theoretical studies have explained the connection between information scrambling in circuits and BPs\cite{sackAvoidingBarrenPlateaus2022}.  We aim to support our analysis of avoiding BPs in MBL circuits by investigating information scrambling. We investigated the entropy growth and information scrambling of the aforementioned model in the MBL region and the thermal region, with the results plotted in Fig.~\ref{fig:otoc} and Fig.~\ref{entropygrowth}. We use the Out-Of-Time-Ordered Correlator(OTOC) to calibrate the system's information scarmbing\cite{fanOutofTimeOrderCorrelationManyBody2017}. More details about OTOC in Appendix. In Fig.~\ref{fig:otoc}, in both PM and DTC regions, as the number of depths growth, the OTOC is decay vary slow, but in thermal region, the OTOC is decay so fast. According to \cite{sackAvoidingBarrenPlateaus2022}, in the circuit structure where BP occurs, the scrambling lightcone eventually extends to the full system, whereas in the circuit structure of the MBL region, information scrambling remains localized even at very deep depths. 

In Fig.~\ref{entropygrowth},  we investigated the bipartite von Neumann entropy of the system with different system size. We calculated the entropy of the system from one half to (qubit-$1$ to qubit-$N/2$) subsystem the other half (qubit-$N/2$ to qubit-$N$) subsystem. In Fig.~\ref{entropygrowth} (a) We can see that in MBL region the entropy growth with the number of the circuit dpeth growth is area law, as the number of qubits increases, the rate of entropy growth with respect to the number of depths remains the same.  In Fig.~\ref{entropygrowth} (b), in the thermal region, the entropy growth with the number of the dpeth growth is volume law. 

\subsection{Details for Optimizer Dynamical}\label{optimizer}
In this subsection, we will investigate the entropy dynamics and cost function dynamics of the MBL Ansatz under different values. We will examine the numerical simulation details of these dynamics and investigate the entropy growth under the MBL unitary. Some earlier theoretical work has attempted to elucidate the relationship between entropy growth and BPs\cite{sackAvoidingBarrenPlateaus2022,Kim_2022,Kim_2022_01,Patti_2021}. We will use our numerical results to analyze and substantiate these relationships.

  The extensive parameter range of the MBL circuit brings effectiveness in combating noise and parameter updates. Using it as the initial circuit in the VQE algorithm can effectively enhance the algorithm. We numerically simulated MBL initial circuit to solve the cost function in the ground state problem for Eq.~(\ref{xxz}), as well as the dynamic evolution of entropy with iteration count. We found that this initial configuration is highly effective in solving many-body ground state problems. 
    Consider the Hamilton in Eq.~(\ref{xxz}), we chose the zero product state $|\boldsymbol{0}\rangle = |00\dots0\rangle$ as the initial state, thus the cost function denote by:
    \begin{equation}
    E(\boldsymbol{J},\boldsymbol{h}, \boldsymbol{g}) = \langle\boldsymbol{0}|\prod_d^D U_F^\dagger(\boldsymbol{J},\boldsymbol{h}, \boldsymbol{g}) H \prod_d^D U_F(\boldsymbol{J},\boldsymbol{h}, \boldsymbol{g})|\boldsymbol{0}\rangle
    \end{equation}
    the gradient direction protocol for parameters $\boldsymbol{\theta} = (\boldsymbol{J},\boldsymbol{h},\boldsymbol{g})$ update is
    \begin{equation}\label{gd}
    \boldsymbol{\theta}^{t+1}=\boldsymbol{\theta}^t-\eta \nabla_{\boldsymbol{\theta}} E(\boldsymbol{\theta}),
    \end{equation}
    where $\eta$ is learning rate and we chose $\eta=0.05$ initially and use the ADAM optimizer to update it. We
use Yao.jl\cite{luoYaoJlExtensible2020}, an extensible and efficient framework to implement our method in classical numerical simulation.

\begin{figure*}[htbp]
    \centering
    \begin{minipage}[b]{0.49\textwidth}
        \centering
        \includegraphics[width=\textwidth]{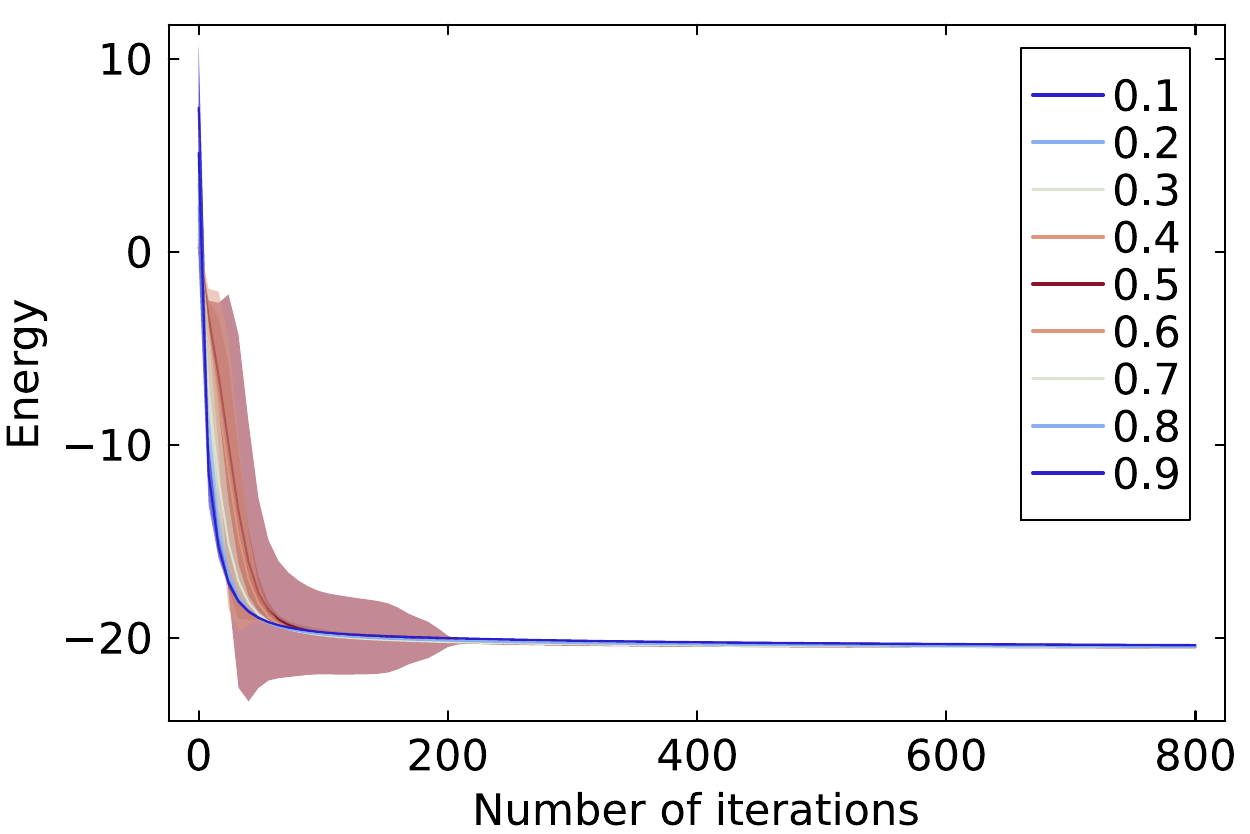}%
        \begin{tikzpicture}[remember picture, overlay]
            \node[anchor=north east, font=\footnotesize, inner sep=0pt] at (-6,4.8) {(a)};
        \end{tikzpicture}
     \end{minipage}
    \hfill
    \begin{minipage}[b]{0.49\textwidth}
        \centering
        \includegraphics[width=\textwidth]{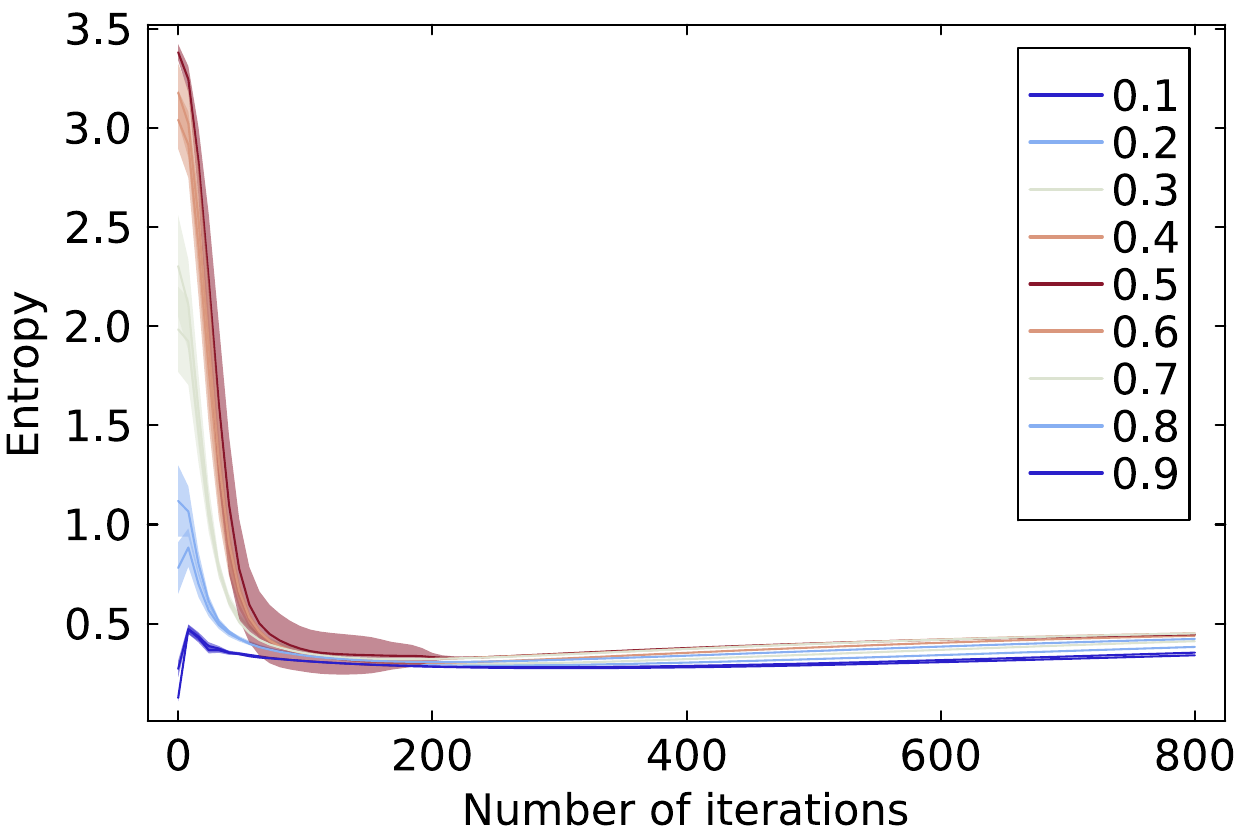}%
        \begin{tikzpicture}[remember picture, overlay]
            \node[anchor=north east, font=\footnotesize, inner sep=0pt] at (-6,4.8) {(b)};
        \end{tikzpicture}
    \end{minipage}
    \caption{Optimizer dynamicals for (a) Cost Function and (b) Entropy, $N = 12$, $D = 2\times N$, across 100 instances to obtain the average values. The shaded area represents the variance relative to the average value of 100 instances at the current iteration step.}
    \label{fig:dynamical}
\end{figure*}

In Fig. \ref{fig:dynamical}, we display our numerical result in the cost function and entropy optimizer dynamical.
We chose the number of qubits as \( N=12 \) and conducted calculations across 100 random instances to obtain the average values. The shaded area illustrates the the variance relative to the average value of 100 instances at the current iteration step. It can be seen that when initialized in the MBL regions, as shown in Fig.~\ref{fig:dynamical} (a), the variance in the thermal region is larger than in the MBL region, which makes sense. This suggests that during fewer iterations, the thermal region searches within a much larger parameter space, leading to greater fluctuations in the cost function and a higher likelihood of encountering BPs. In contrast, the MBL region searches within a smaller space closer to the ground state, exhibiting better stability.
From Fig.~\ref{fig:dynamical} (b), it is evident that the MBL region starts from a very low entropy, which is closer to the characteristics of the ground state of the Hamiltonian we are seeking. This results in a larger overlap with the target and a smaller search space for the states. On the other hand, the thermal circuit starts from a much higher entropy, leading to a larger state space to search, and therefore, the MBL initialization is more efficient. 

\section{Discussion and Conclusion}\label{sec:discussion}

Our research has yielded significant findings that contribute to the advancement of the VQE algorithm through the incorporation of MBL principles. Firstly, we have demonstrated that the circuit structure within the MBL regime can achieve substantial gradients, effectively circumventing the BP issue that plagues traditional VQE approaches. This discovery is pivotal as it addresses a critical challenge in scaling quantum algorithms across an increasing number of qubits.

Secondly, we presented an in-depth analysis of the ansatz's entropy dynamics and cost function dynamics across various initialization values. This comprehensive exploration allowed us to visualize the intricate interplay between the initialization parameters and the system's performance, providing valuable insights into how different starting points can influence the convergence and efficiency of the VQE algorithm.

Furthermore, by calculating the entropy growth within the MBL circuit, we unveiled a connection between the BP phenomenon and the volume law typically associated with chaotic circuits. This revelation is particularly noteworthy as it offers a fresh perspective on the relationship between entanglement and the emergence of BPs, potentially guiding future research in quantum algorithm design.

\begin{figure}[htpb]
    \centering
    \includegraphics[width=0.45\textwidth]{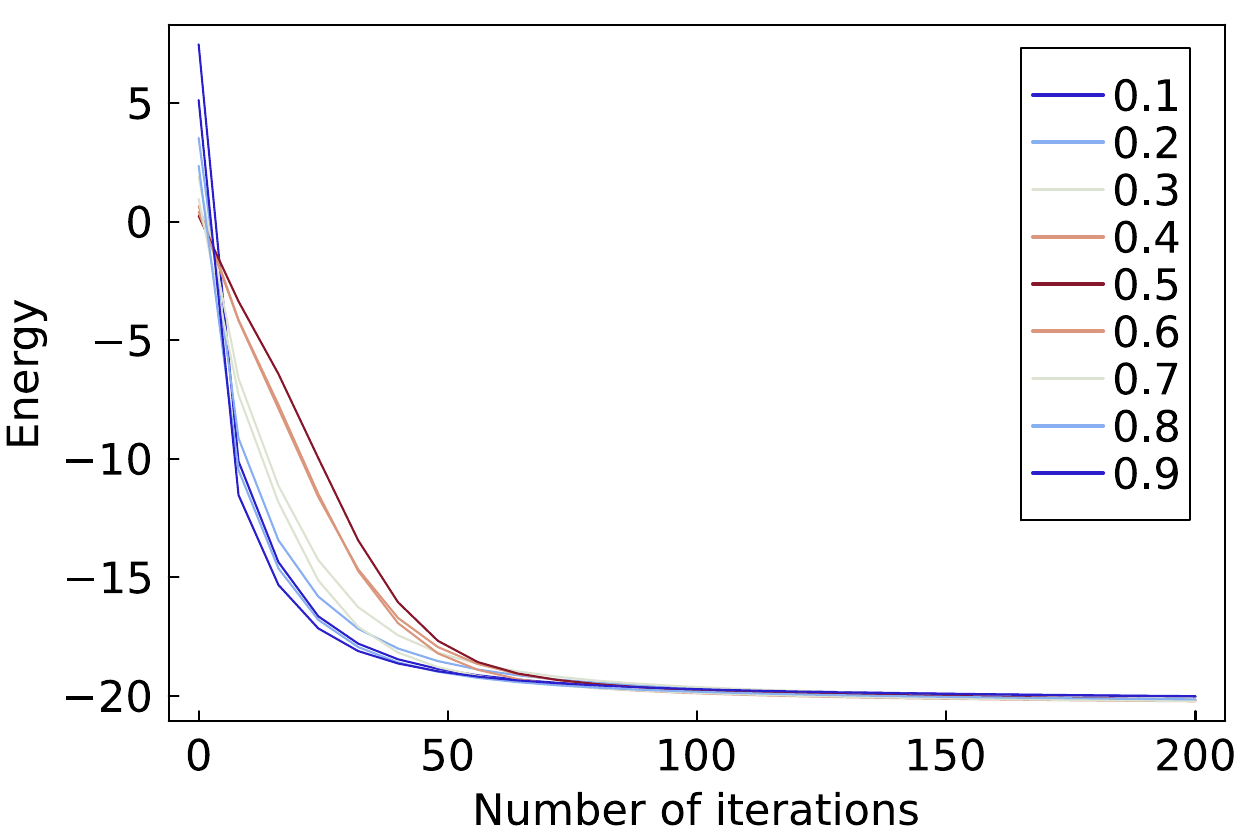}
    \caption{Energy optimizer dynamical, across 1000 instances and set $N = 12, D = 2 \times N$. After about sixty iterations, the local minima in the MBL (Many Body Localization) region are larger than those in the thermal region, with the thermal region being able to reach smaller local minima.}
    \label{fig:argue}
\end{figure}

Lastly, our dynamical analysis revealed that initial parameter values within the MBL range can significantly enhance the efficiency of the VQE algorithm. However, we also observed that beyond a certain number of iterations, the performance of MBL-optimized parameters may not surpass those of thermally initialized values as shown in Fig.~\ref{fig:argue}. This intriguing finding opens up new avenues for exploration, particularly in understanding the long-term behavior of quantum algorithms and identifying optimal strategies for maintaining high-gradient regions throughout the optimization process. Furthermore, recent research has demonstrated long-lived topological time-crystalline order on a 2D quantum processor\cite{xiang2024}. Our study shows the effectiveness of 1D time crystal structures in solving 1D Hamiltonians. We expect that in more complex 2D systems, the corresponding structures could also improve 2D VQE. We speculate that the characteristics shown in Fig.~\ref{fig:argue} are related to the topological time-crystalline order of non-equilibrium states and the topological order of many-body ground states under long-term evolution. We hope that further research with 2D systems can clarify these issues.

In summary, our work not only provides a robust strategy for mitigating the BP problem in VQE but also deepens our understanding of the underlying dynamics that govern quantum algorithm performance. The insights gained from this study lay the groundwork for further refinements in quantum optimization techniques and the development of more resilient quantum computing applications.
In addition to our primary findings, we would like to delve into further discussions that could enrich our understanding of the MBL-VQE approach and its implications in quantum computing.

Some literature has explored the BP problem in periodically driven random parameterized circuits using techniques inspired by quantum optimal control(QOC)\cite{liuMitigatingBarrenPlateaus2022,
laroccaDiagnosingBarrenPlateaus2022, fontana2024adjoint}. From a QOC standpoint, we argue that MBL-structured circuits can also construct the corresponding Lie algebra, leading to a Dynamical Lie Algebra. Compared to the literature, such circuit structures should represent an uncontrollable, irreducible case. However, to the best of our knowledge, no studies have applied QOC methods to finite-size MBL. This presents an intriguing area for future research, as it could offer novel insights into the controllability of quantum systems within the MBL regione.

Our numerical results suggest that after a certain number of iterations, the parameter range of the circuit may deviate from the MBL region. In some cases, the performance of the circuit may not be as favorable as that of general hardware-efficient circuits. We hypothesize that such deviations might be associated with phenomena like many-body resonances and avalanches\cite{morningstarAvalanchesManybodyResonances2022}. These complex dynamics could play a crucial role in the long-term behavior of quantum circuits and may require advanced techniques to manage and mitigate their effects on the optimization process.

Random Matrix Theory(RMT) is a powerful tool commonly used to study MBL. According to RMT, thermal circuits are generally believed to follow Gaussian Unitary Ensemble (GUE) or Gaussian Orthogonal Ensemble (GOE)  level ratios, while MBL circuits adhere to Poisson distributions\cite{Buijsman_2019}. Our work demonstrates the effectiveness of MBL circuits in avoiding BPs, which suggests that future research could establish a more precise connection between these ensembles and unitary t-designs. Such a connection could pave the way for a new paradigm in the study of random circuit variational algorithms, offering a deeper theoretical foundation for the design and optimization of quantum algorithms.

Our research opens up several avenues for future exploration. The application of QOC methods to MBL, understanding the role of many-body dynamics in circuit performance, and leveraging RMT to refine our approach to random circuit variational algorithms are all promising directions. These discussions not only complement our findings but also contribute to the broader discourse on quantum algorithm optimization and the development of robust quantum computing technologies.

\textit{Add Note:} When we finished this article we noticed that there were two articles\cite{park2024,cao2024} that did similar work. In comparison, our work not only analyzes the parameter conditions of the MBL circuit but also provides a detailed study of the dynamics of the system during the optimization process. Through this approach, we have been able to reveal key dynamical characteristics when optimizing quantum algorithms in the MBL phase, an area that has not been explored in previous research.

\section*{Acknowledgements} \label{sec:acknowledgements}
We extend our sincere gratitude to Jingze Zhuang, Yiming Huang, Qiming Ding and Zixuan Huo for their invaluable discussions and insights throughout the course of this research.  This work is supported by Beijing Institute of Technology Research Fund Program for Young Scholars.

\bibliographystyle{naturemag}
\bibliography{refs}

\appendix
\begin{widetext}
\section{Entropy and information scrambling} \label{app}
\setcounter{equation}{0} 
\renewcommand{\theequation}{A\arabic{equation}}
\renewcommand{\thefigure}{A\arabic{figure}}
\setcounter{figure}{0}
In quantum information and statistical mechanics, the area law and volume law for entropy increase describe the behavior of entanglement entropy in different physical states. The area law states that, for a quantum system, the entanglement entropy is proportional to the area of the subsystem's boundary $S_A \propto |\partial A|$, typically applying to low-energy states such as the ground state and some low-lying excited states. In contrast, the volume law states that, for certain highly excited states, the entanglement entropy is proportional to the volume of the subsystem $S_A \propto V_A$, which generally applies to thermalized systems and non-equilibrium states like highly energy excited states. In these cases, every part of the system is highly entangled with others. Understanding these two laws is crucial for studying quantum many-body systems and quantum information. However, for MBL systems, highly excited states still obey the area law.

    As a common method for tracking information scrambling, the out-of-time-order correlator (OTOC) has drawn a lot of attention in both the gravity physics, the condensed matter physics and quantum information.  We numerically simulated the following OTOC:
    \begin{equation}\label{infotoc}
        F(t)=\frac{1}{2^D} \sum_n\left\langle n\left|\hat{U}^{\dagger} \hat{\tau}_i \hat{U} \hat{\tau}_j\hat{U}^{\dagger} \hat{\tau}_i \hat{U} \hat{\tau}_j\right| n\right\rangle
    \end{equation}

    This corresponds to calculating the overlap of two time-evolved operators in the Heisenberg picture for a system initially in an infinite temperature thermal state. This simulation can be easily implemented using numerical methods. We chose
    $\hat{\tau}_i$ and $\hat{\tau}_j$ as Pauli operator $\sigma^x, \sigma^y$ and $\sigma^z$. A more general form of OTOC is 
    \begin{equation}
    \label{gotoc}
    F(t) = \text{Tr}[\hat{U}^{\dagger} \hat{\tau}_i \hat{U} \hat{\tau}_j\hat{U}^{\dagger} \hat{\tau}_i \hat{U} \hat{\tau}_j\rho]
    \end{equation} 
    
    where $\rho$ can be pure state or mix state, Eq.~(\ref{infotoc}) is just chosen $\rho = \frac{1}{2^D} \hat{I}$ in Eq.~(\ref{gotoc}). 
    In the main text, we choose
$i$ to be the first qubit and $j$ to be the last qubit. We take $\hat{\tau}_i$ and $\hat{\tau}_j$ to both be the 
Pauli X gate $\sigma^x$.

\end{widetext}

\end{document}

%% file: preamble.tex
\usepackage{amsthm}
\usepackage{mathtools}
\usepackage{physics}
\usepackage{xcolor}
\usepackage{graphicx}
\usepackage[left=23mm,right=13mm,top=35mm,columnsep=15pt]{geometry} 
\usepackage{adjustbox}
\usepackage{placeins}
\usepackage[T1]{fontenc}
\usepackage{lipsum}
\usepackage{csquotes}